\documentclass{aa}
\usepackage{psfig}

\def\Ni{\noindent}
\def\msun{$M_{\odot}$}
\def\etal{{\it et al.}}
\def\ros{{\it ROSAT}}

\def\fss{\hbox{$.\!\!^{\rm s}$}}        % Fractions of seconds
\def\farcm{\hbox{$.\mkern-4mu^\prime$}}  % Fractions of arcmin
\def\asec{\ifmmode ^{\prime\prime}\else$^{\prime\prime}$\fi}
\def\amin{\ifmmode ^{\prime}\else$^{\prime}$\fi}
\def\D{\ifmmode ^{\circ}\else$^{\circ}$\fi}
\def\H{$^{\rm h}$}\def\M{$^{\rm m}$}
\newbox\grsign \setbox\grsign=\hbox{$>$}
\newdimen\grdimen \grdimen=\ht\grsign
\newbox\laxbox \newbox\gaxbox
\setbox\gaxbox=\hbox{\raise.5ex\hbox{$>$}\llap
     {\lower.5ex\hbox{$\sim$}}}\ht1=\grdimen\dp1=0pt
\setbox\laxbox=\hbox{\raise.5ex\hbox{$<$}\llap
     {\lower.5ex\hbox{$\sim$}}}\ht2=\grdimen\dp2=0pt
\def\gax{\mathrel{\copy\gaxbox}}
\def\lax{\mathrel{\copy\laxbox}}
\def\mdot{$\dot M$}
\def\msun{$M_{\odot}$}
\font\pr=cmr7
\def\II{{\pr II}}
\def\III{{\pr III}}
\def\IV{{\pr IV}}
\def\V{{\pr V}}

\begin{document}

   \thesaurus{06         % A&A Section 6: Form. struct. and evolut. of stars
               (02.01.2 % Accretion, accretion disks
                08.02.1 % Stars: binaries: close
                08.09.2 % Stars: individual
                08.14.2 % Stars: novae, cataclysmic variables
                13.25.5)} % X-rays: stars

\title{Soft X-ray emission of VY Scl stars during optical high state}

\author{J. Greiner}
\institute{Astropysikalisches Institut Potsdam, An der Sternwarte 16, 
14482 Potsdam, Germany}

\offprints{J. Greiner; jgreiner@aip.de}

   \date{Received  27 April 1998; accepted 14 May 1998}

   \maketitle

\begin{abstract}
I have collected all available ROSAT observations of VY Scl stars including
those during the all-sky survey to investigate the presence, strength
and spectrum of soft X-ray emission (0.1--2.4 keV) of this group of nova-like
variables. A total of 9 out of the 14 VY Scl stars are detected with 
ROSAT, mostly during optical high states. Interestingly, all detections
during the optical high state have very similar X-ray spectral characteristics.
I find that a simple blackbody model gives a reasonably good fit in all 
cases, with temperatures falling in the narrow range between 0.25--0.5 keV.
Possible emission mechanisms are discussed.
\end{abstract}

\vspace{-0.4cm}


\section{Introduction}

VY Scl stars are a subclass of nova-like, cataclysmic variables which are 
bright most of the time, but occasionally drop in brightness at irregular 
intervals (e.g. Warner 1995). The transitions between the brightness levels
occur on a time scale of days to weeks. These variables are 
typically found at $P >$ 3 hrs and with large mass transfer
rate \mdot\ (upper right corner of the $P_{\rm orb}-$\mdot\ diagram of
Osaki 1996), and thus are thought to be steady accretors (or dwarf novae in a 
state of continuous eruption as suggested by Kraft 1964) with hot disks.
Their disks are thus assumed to be thermally and tidally stable.

Evidence for high \mdot\ is based on measures of the absolute magnitude
of VY Scl stars during their high state. The accretion disks then are assumed
to be optically thick and can be approximated by models of quasi-steady disks
(see e.g. Warner 1987 for a summary). Absolute magnitudes can be determined 
in various ways, and range between $M_{\rm V}$ = 3--6 mag for VY Scl stars
(Warner 1987, see also Tab. \ref{xlist}),
corresponding to mass transfer rates of up to 5$\times$10$^{-8}$ \msun/yr.

An interesting, partly overlapping group of high \mdot\ cataclysmic systems  
are the so-called SW Sex stars (three of the five SW Sex stars belong
to the VY Scl group) which all are eclipsing systems
but show single-peaked emission lines remaining largely unobscured
during primary eclipse. This has been explained in terms of a combination
of an accretion stream which  is overflowing the initial impact onto the disk
with the effect of a strong accretion disk wind (e.g. Hellier 1996). 
The observations  of these properties in 
eclipsing systems only is certainly a selection bias, and it remains to be seen
whether other VY Scl stars also exhibit some or all of the properties which
presently make up the SW Sex classification criteria.

Based on the fact that VY Scl stars have similar low states like AM Her 
binaries which have no disks, low states are thought to involve drops in the
mass transfer rate from the secondary.
Livio \& Pringle (1994) proposed a model for the 
optical brightness drops of  VY Scl stars in which
the reduced mass transfer rate is caused by a magnetic spot covering 
temporarily the $L_1$ region. This mechanism works predominantly at short 
orbital periods because the level of magnetic activity increases with the 
rotation rate of the star (which in turn is coupled to the orbit).
The same idea has recently been expanded and applied to detailed disk 
instability modelling (King \& Cannizzo 1998). It was shown that a
simple reduction in mass transfer rate from the secondary is not
sufficient because then, after the transition of the disk to a cool state, the
disk should show outbursts like in dwarf novae. Since such outbursts have
never been observed from VY Scl stars during optical low-state,
King \& Cannizzo (1998) concluded that all disk mass must be drained away 
after the transition of the system into the optical low state.

\begin{table*}
\caption{Compilation of basic properties of the 14 presently known
      VY Scl stars$^{(1,2)}$}
\begin{tabular}{cccrccccccc}
\hline
\noalign{\smallskip}
    Name~ & R.A. & Decl. & b$^{\rm II}$ & Mag.  & 
   P$_{\rm orb}$ & d & $\!\!$M$_{\rm V}^{\rm low}$/M$_{\rm V}^{\rm high}\!\!$& 
      i$^{\rm (3)}$ &    M$_{\rm WD}$ & M$_{\rm D}$  \\
         & (2000.0)   & (2000.0) &    & range   &  
    (min) &       (pc) &  (mag)      & (\D) & (\msun) & (\msun)   \\ 
\noalign{\smallskip}
\hline
\noalign{\smallskip}
 PX And & 00\H30\M05\fss9 & $\!\!$+26\D17\amin26\asec & --36  & 14.8--17.0
     & 211 & $>$180 & $<$9/$<$7 & $\approx$74$^{\rm ecl}$ & 0.2$^{(4)}$ 
     & 0.32    \\
 TT Ari & 02\H06\M53\fss1 & $\!\!$+15\D17\amin42\asec & --43 & ~\,9.5--14.5
     & 198 & $\!$180--200$\!$ & 9/4 &30--40 &$\approx$1 & 0.35  \\
 KR Aur & 06\H15\M44\fss0 & $\!\!$+28\D35\amin08\asec & +6 & 11.3--19.0
     &  234 & 180 & 12/5 & $\lax$40 & 0.7 & 0.48  \\
 BZ Cam & 06\H29\M34\fss1 & $\!\!$+71\D04\amin36\asec & --24 & 12.5--14.1
     & 221 & $\approx$500$^{(5)}$ & 6/4$^{(5)}$  & $\lax$40 & 0.1--1.0 
     & 0.3--0.35  \\
 BH Lyn & 08\H22\M36\fss1 & $\!\!$+51\D05\amin24\asec &  +35 & 13.7--17.2
     & 224  &  &   & $\approx$79$^{\rm ecl}$ & 0.37--1.4 & 0.22--0.5  \\
 DW UMa  & 10\H33\M53\fss1 & $\!\!$+58\D46\amin54\asec & +50 & 13.8--18.1
     & 197 & $\approx$850 & 7.5/3 &$>$72$^{\rm ecl}$ & 0.15--0.6 & 0.15--0.3 \\
 LX Ser & 15\H38\M00\fss2 & $\!\!$+18\D52\amin02\asec &  +51 & 13.3--17.4
     & 228 & $\!$250--460$\!$ & 9/5 & 75$^{\rm ecl}$ & 0.32--0.48 &0.32--0.39\\
$\!$V442 Oph & 17\H32\M15\fss2 & $\!$--16\D15\amin23\asec & +9 & 12.6--15.5
     &  202  & $>$80 & $<$10/$<$7  & $\lax$65 & 0.35--0.45 & 0.3--0.37  \\
 MV Lyr & 19\H07\M16\fss4 & $\!\!$+44\D01\amin07\asec & +16 & 12.2--18.0
     &  193  & 320 & 10.2/6.0 & 9--14 & 0.4--1.4 & 0.17 \\
 V794  Aql & 20\H17\M34\fss0 & $\!$--03\D39\amin52\asec & --21  & 13.7--20.2 
     & $\!$240-330$\!$ & 200 & 14/7  & 22--56 & 0.5--1.2 & 0.46--0.6  \\
$\!$V751 Cyg & 20\H52\M12\fss9 & $\!\!$+44\D19\amin25\asec & --0 & 13.2--17.8
     & $\approx$360    & 430 & 8.2/3.6  & & &  \\
$\!$V425 Cas & 23\H03\M46\fss7 & $\!\!$+53\D17\amin14\asec & --6 & 14.5--18.5
     & 216      &  &      & 16--34 & 0.55--1.2 & 0.29--0.33  \\
 VY    Scl & 23\H29\M00\fss5 & $\!$--29\D46\amin47\asec & --72 & 12.9--18.5
     & 239 & $\approx$500 &  9/4   & 25--40 & 0.8--1.4 & 0.23--0.42  \\
  VZ    Scl & 23\H50\M09\fss2 & $\!$--26\D22\amin53\asec & --76 & 15.6--20.0
     & 208 & 530 & 11.4/3  & $>$76$^{\rm ecl}$ & 0.3--1.0 & 0.32 \\
\hline
\noalign{\smallskip}
\end{tabular}

\noindent{\Ni 
   $^{(1)}$ The coordinates and most of the optical magnitudes are taken from 
            Downes \& Shara (1993). Note that
            the Simbad coordinates are sometimes less accurate. \\
   $^{(2)}$ References for table values:
            PX And: Thorstensen \etal\ 1991, Still \etal\ 1995,
            TT Ari: Cowley \etal\ 1975, Shafter \etal\ 1985,
            KR Aur: Shafter 1983a, Antov \etal\ 1996,
            BZ Cam: Lu \& Hutchings 1985, Krautter \etal\ 1987,
            BH Lyn: Richter 1989, Andronov \etal\ 1989, Dhillon \etal\ 1992, 
                    Hoard \& Szkody 1997,
            DW UMa: Hessmann 1990, Honeycutt \etal\ 1993, Dhillon \etal\ 1994,
            LX Ser: Young \etal\ 1981, Eason \etal\ 1984, 
            V442 Oph: Shafter \& Ulrich 1982, Szkody \& Shafter 1983,
            MV Lyr: Schneider \etal\ 1981,
            V794 Aql: Shafter 1983b,c,  Honeycutt \& Schlegel 1985,
            V751 Cyg: Robinson \etal\ 1974, Bell \& Walker 1980, 
                      Greiner \etal\ 1998,
            V425 Cas: Shafter \& Ulrich 1982, Shafter 1983c, Wenzel 1987, 
            VY Scl: Hutchings \& Cowley 1984,
            VZ Scl: Schaefer \& Patterson 1982, Sherrington \etal\ 1984, 
                    O'Donoghue \etal\ 1987 \\
   $^{(3)}$ Eclipsing systems are marked by ``ecl''. \\
   $^{(4)}$ The formal best-fit result of M$_{\rm WD} \approx$0.2 \msun\ has
            been regarded as implausible because it implies mass transfer 
            on a dynamical time scale (Thorstensen \etal\ 1991). \\
   $^{(5)}$ M$_{\rm V}$=4 has been assumed in deriving the distance 
           (Krautter \etal\ 1987).
     }
\label{xlist}
\end{table*}

Previously (pre-ROSAT) known X-ray emission from VY Scl stars include
V794 Aql (Szkody \etal\ 1981), 
TT Ari during the optical high state (Jensen \etal\ 1983), 
KR Aur during the optical high state (Mufson \etal\ 1980, Singh \etal\ 1993), 
LX Ser (Szkody 1981).
These investigations have consistently concluded that the  high X-ray 
luminosity expected from the boundary layer ($\sim$10$^{34}-10^{35}$ erg/s),
based on the high 
M$_{\rm V}$ and thus accretion rate of the order of 10$^{-8}$ \msun/yr)
was not detectable.

In order to relate the findings of this X-ray survey of VY Scl stars to
some possibly underlying physical quantities, I have compiled some
important system parameters like the apparent magnitude range, orbital
period, distance, inclination and mass of the binary components from the 
available literature (Tab. \ref{xlist}).
Also, I have collected the brightness estimates of the VY Scl stars over the
last seven years from various sources in order to determine the optical state 
during which the ROSAT observations have been performed (Fig. \ref{lc}). 
Many of the VY Scl stars are monitored by various amateur astronomers 
around the world and thus a substantial amount of monitoring data was
available from the AAVSO, AFOEV and VSNET databases.

In the following I present a complete overview of the ROSAT observations
of VY Scl stars during the all-sky survey as well as in subsequent
pointed observations.
Results for some of the ROSAT observations have been reported already earlier:
on MV Lyr and KR Aur during their optical high state (Schlegel \& Singh 
1995; Richman 1996), on TT Ari (Baykal \etal\ 1995; Richman 1996), 
on BZ Cam (van Teeseling \& Verbunt 1994), and on VZ Scl and DW UMa  
(van Teeseling \etal\ 1996). 
The results of the \ros\ all-sky survey detections of
V794 Aql and BZ Cam have already been mentioned  in Verbunt \etal\ (1997).

\begin{table*}
\vspace{-0.15cm}
\caption{ROSAT observations of VY Scl stars$^{(1)}$}
\begin{tabular}{cccccccccr}
\hline
\noalign{\smallskip}
 Name & Date  &  T$_{\rm exp}$ & offaxis  &  CR$^{(2)}$ & HR1 & HR2 
                        & log L$_{\rm X}^{(3)}$     & opt.  & D$^{(4)}$ \\
      &      & (sec) & angle & (cts/s) & & 
                       & (erg/s) &  $\!\!$state$\!\!$ &  \\
\noalign{\smallskip}
\hline
\noalign{\smallskip}
 PX And & $\!\!$Dec. 31, 1990--Jan. 1, 1991 & ~\,350 & 0--55\amin & $<$0.022 
                                             & -- & -- & --$^{(5)}$ & & -- \\
        & Jul. 4/5, 1991  & $\!\!\!$26\,294 & 0\farcm3 & 0.0052$\pm$0.0005 &
          0.79$\pm$0.08 & 0.24$\pm$0.09 & $>$29.7 & & ~\,5\asec \\
\noalign{\smallskip}
 TT Ari & Jan. 20/21, 1991 & ~~275 & 0--55\amin & 0.360$\pm$0.037  & 
          0.80$\pm$0.06 & 0.34$\pm$0.09 & 31.4 & high & ~\,4\asec \\
        & Aug. 1/2, 1991 & $\!\!\!$24\,465 & 0\farcm3 & 0.366$\pm$0.006 &
          0.82$\pm$0.01 & 0.34$\pm$0.01 & 31.4 & high & 1\asec \\
\noalign{\smallskip}
 KR Aur & Sep. 14/15, 1990 & ~~360 & 0--55\amin & 0.065$\pm$0.014  & 
          1.00$\pm$0.00 & 0.75$\pm$0.14 & 30.7 & high & ~\,7\asec \\
   & Sep. 28--Oct. 6, 1992  & $\!\!\!$17\,255 & 0\farcm3 & 0.064$\pm$0.002 &
        0.94$\pm$0.01 & 0.52$\pm$0.03 & 30.7 & high & ~\,2\asec \\
\noalign{\smallskip}
 BZ Cam & Sep. 14/16, 1990 &  ~~430 & 0--55\amin & 0.077$\pm$0.014  & 
          1.00$\pm$0.00 & 0.17$\pm$0.18 & 31.9 & & 17\asec \\
        & Sep. 29, 1992  & 6\,117 & 0\farcm2 & 0.074$\pm$0.004 &
          0.90$\pm$0.02 & 0.30$\pm$0.05 & 31.8 & high & ~\,4\asec \\
        & Sep. 3, 1993  & 4\,591 & 0\farcm2 & 0.062$\pm$0.004 &
          0.89$\pm$0.03 & 0.37$\pm$0.06 & 31.6 & & ~\,4\asec \\
\noalign{\smallskip}
 BH Lyn & Oct. 7--9, 1990 & ~~460 & 0--55\amin & $<$0.006 & -- & -- & 
               --$^{(6)}$ & & -- \\
\noalign{\smallskip}
 DW UMa & Oct. 25--28, 1990 &  ~~370 & 0--55\amin & $<$0.033 & -- & -- 
          & & high & -- \\
        & Oct. 15, 1992  & 3\,395 & 0\farcm1 & 0.011$\pm$0.002 &
          0.28$\pm$0.14 & 0.31$\pm$0.16 & 31.3$^{(7)}$ & rise & ~\,4\asec \\
        & Oct. 15, 1992  & 1\,710 & 0\farcm1 & 0.013$\pm$0.003 &
          0.42$\pm$0.20 & 0.44$\pm$0.20 & 31.2$^{(7)}$ & rise & ~\,6\asec \\
\noalign{\smallskip}
 LX Ser & Aug. 8--10, 1991 & ~~345 & 0--55\amin & $<$0.037 & -- & -- 
           & $<$31.1$^{(7)}$  & high & -- \\
\noalign{\smallskip}
 V442 Oph & Sep. 4/5,   1990  & ~~315 & 0--55\amin & $<$0.014 & -- 
              & -- &  --$^{(5)}$  &  & -- \\
          & Sep. 22/23, 1992  & $\!\!\!$11\,385 & 41 & $<$0.0064 & --
              & -- & --$^{(5)}$ & & -- \\
\noalign{\smallskip}
 MV Lyr & Oct. 11--15, 1990 &  ~~722 & 0--55\amin & 0.079$\pm$0.011  & 
            0.82$\pm$0.09 & 0.69$\pm$0.10 & 31.3 & high & ~\,6\asec \\
        & Nov. 4--8, 1992  & $\!\!\!$20\,250 & 0\farcm2 & 0.069$\pm$0.002 &
            0.84$\pm$0.02 & 0.45$\pm$0.02 & 31.2 & high  & 4\asec \\
        & May 28, 1996  & 2\,218 & 0\farcm1 & $<$0.0008$^{(1)}$ &
         --  & -- & $<$29.7 & low & -- \\
\noalign{\smallskip}
 V794 Aql & Oct. 18, 1990 &  ~~240 & 0--55\amin & 0.093$\pm$0.021  &
           0.87$\pm$0.11 & 0.44$\pm$0.19 & 30.8$^{(7)}$ & high & 11\asec \\
\noalign{\smallskip}
 V751 Cyg & Nov. 19/20, 1990  & ~~370 & 0--55\amin & $<$0.019 & -- & -- 
                                               &$<$30.3$^{(7)}$ & high & -- \\
          & Nov. 3, 1992   &  3637 & 52     & $<$0.0058 & --
             & --  & $<$30.8$^{(7)}$ & high & -- \\
          & Jun. 3, 1997   &  4663 & 0\farcm3 & 0.11$\pm$0.02$^{(1)}$ 
             & -- & -- & --$^{(1)}$ & low & 2\asec \\
          & Dec. 2--8, 1997   & 10813 & 0\farcm2 & 0.08$\pm$0.02$^{(1)}$
             & -- & -- & --$^{(1)}$ & low & 7\asec \\
\noalign{\smallskip}
 V425 Cas & Dec. 30/31, 1990 & ~~380 & 0--55\amin & $<$0.019 & -- & -- 
     & --$^{(6)}$ & high & -- \\
\noalign{\smallskip}
 VY Scl & Nov. 23--25, 1990 &  ~~~85 & 0--55\amin & $<$0.13  & -- & -- 
     & $<$31.7$^{(7)}$ & & -- \\
\noalign{\smallskip}
 VZ Scl & Nov. 29--Dec. 1, 1990 & ~~290 & 0--55\amin & $<$0.014 & -- & -- 
     & & high & -- \\
        & Dec. 5, 1992  & 3\,354 & 0\farcm2 & 0.004$\pm$0.001 &
            0.35$\pm$0.25 & 0.03$\pm$0.27 & 30.3$^{(7)}$ & high & ~\,7\asec \\
        & Dec. 5, 1992 &  2\,237 & 0\farcm2 & 0.006$\pm$0.002 &
            0.70$\pm$0.24 & 0.12$\pm$0.29 & 30.5$^{(7)}$ & high & ~\,6\asec \\
\noalign{\smallskip}
\hline
\end{tabular}

\noindent{\Ni\small 
 $^{(1)}$  All observations except the May 1996 pointing on MV Lyr and the
           1997 pointings on V751 Cyg have been performed with the ROSAT
           PSPC. For MV Lyr the HRI count rate has been transformed into a 
           PSPC rate by using the spectral fit parameters of the PSPC 
           observation. For V751 Cyg see Greiner \etal\ (1998). \\
 $^{(2)}$  Count rates are calculated for the 0.1--2.4 keV range (= channels
           11-240). Upper limits are 3$\sigma$ confidence level.\\
 $^{(3)}$  The distances of Tab. \ref{xlist} have been used, in particular 
           200 pc for TT Ari and 460 pc for LX Ser. \\
 $^{(4)}$ Distance between best-fit X-ray position and optical position of
       presumed counterpart. \\
 $^{(5)}$ The upper limit of the count rate has not been combined with the
          lower limit in distance. \\
 $^{(6)}$ No distance known. \\
 $^{(7)}$ A temperature of 0.4 keV has been assumed for the conversion of the 
        upper limit count rates or when the number of counts was below 100. } 
\label{xsurvlog}
\vspace*{-0.15cm}
\end{table*}

\section{Observations}

\subsection{ROSAT all-sky survey and pointed observations}

The ROSAT all-sky survey was performed between July 1990 and January 1991
and typically reached 300--400 sec exposure time near the ecliptic equator.
Each position on the sky is scanned for 2 days (at the ecliptic equator)
or more, and bright X-ray sources can be localized to about 10\asec--20\asec\
accuracy. A total of 5 out of the 14 VY Scl stars are detected within the 
ROSAT all-sky survey, and upper limits are derived for the others.  
Tab. \ref{xsurvlog} gives the date and total duration of the exposure, 
the count rate in the position-sensitive proportional counter (PSPC), the
two hardness ratios HR1 and HR2, and the distance between the best-fit X-ray 
position and the nominal optical position.
The hardness ratio HR1 is defined as the normalized count difference 
(N$_{\rm 50-200}$ -- N$_{\rm 10-40}$)/(N$_{\rm 10-40}$ + N$_{\rm 50-200}$), 
where N$_{\rm a-b}$ 
denotes the number of counts in the PSPC between channel a  and channel b.
Similarly, the hardness ratio HR2 is defined as 
(N$_{\rm 91-200}$ -- N$_{\rm 50-90}$)/N$_{\rm 50-200}$
(note that HR1 is sensitive to the absorbing column).

A total of 8 out of 14 VY Scl stars have been the target of pointed
ROSAT PSPC observations, and in all cases there have been positive detections.
In 2 cases (V442 Oph and V751 Cyg), ROSAT PSPC pointings towards other prime 
targets led to serendipituous coverage, but yield only upper limits.

Furthermore, two VY Scl stars (MV Lyr and V751 Cyg) have been observed with
the ROSAT high-resolution imager (HRI) as  target-of-opportunity during
times of optical low states, and results of these observations are
reported elsewhere (Greiner \etal\ 1998).

A summary of all ROSAT observations of the 14 VY Scl stars  is given
in Tab. \ref{xsurvlog}.

%%-----------------------------------------------------------
   \begin{figure*}
    \vbox{\psfig{figure=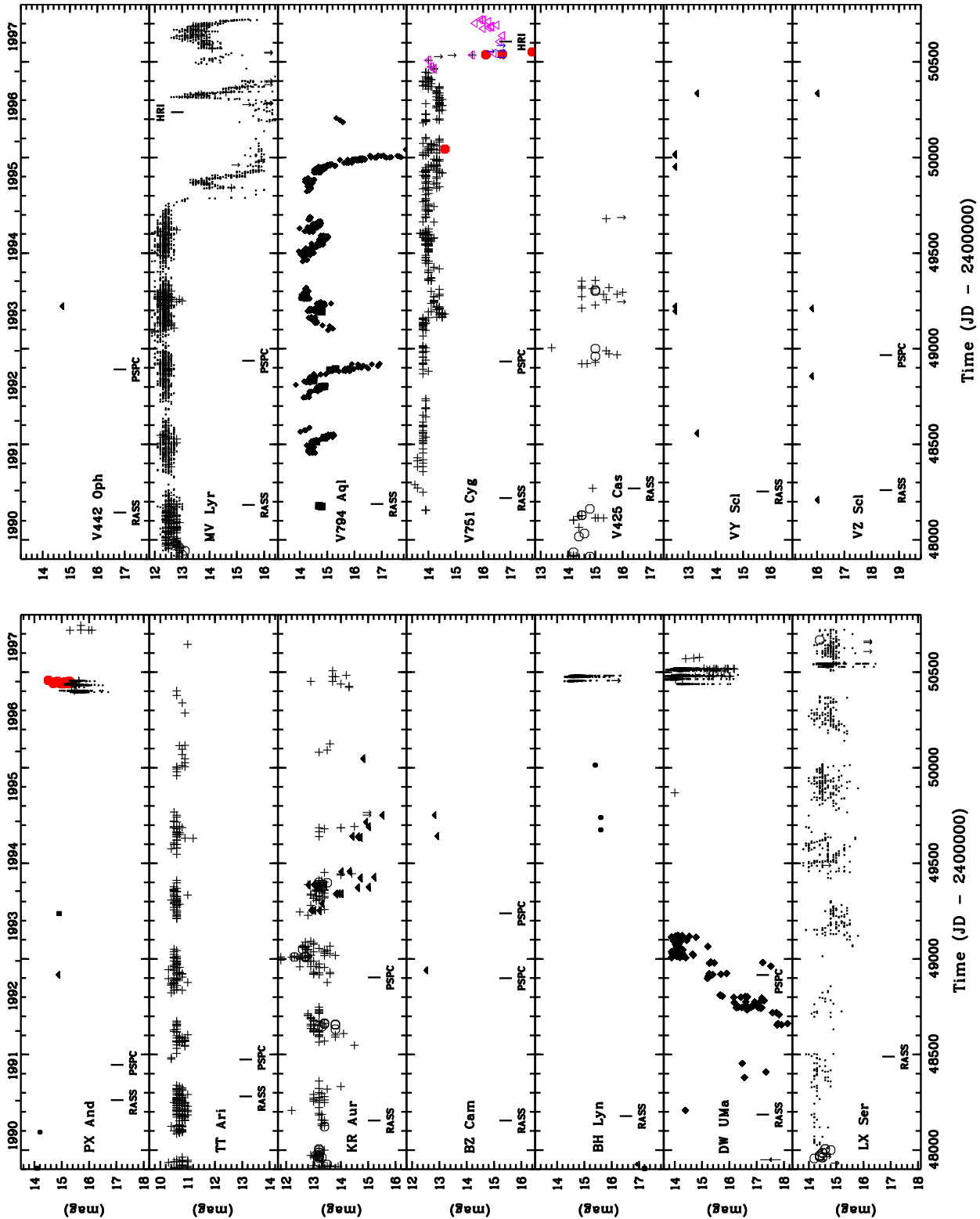,width=17.cm,%
           bbllx=2.3cm,bblly=1.5cm,bburx=20.2cm,bbury=23.5cm,clip=}}\par
    \caption[pha]{Optical light curves of the 14 VY Scl stars over the last 
        7 years based on data from AAVSO, VSNET and AFOEV 
        (scales are identical in all panels):
        crosses denote secure measurements while uncertain brightness 
        estimates are marked by an open circle.
        Arrows denote upper limits.
        In addition, individual observations are included as follows:
      PX And: Thorstensen \etal\ (1991; hexagon), Shakhovskoy \etal\ (1995; 
              triangle), Still \etal\ (1995; squares),
      KR Aur: Antov \etal\ (1996; triangles), 
      BZ Cam: Patterson \etal\ (1996; triangles),
      DW UMa: Hessmann (1990; triangle with large error bar),
              Honeycutt \etal\ (1993; lozenge),
      V442 Oph: UK Schmidt (triangles),
      V794 Aql: Richter (1998; squares), Honeycutt \etal\ (1994; lozenge), 
      VY Scl: UK Schmidt (triangles),
      VZ Scl: UK Schmidt (triangles).
         At the bottom of each panel the times of the ROSAT observations 
        are marked.
          }
      \label{lc}
   \end{figure*}
%%______________________________________________________________

   \begin{table*}
     \caption{X-ray spectral fit results of VY Scl stars with a blackbody 
          model$^{(1)}$}
      \begin{tabular}{crccccccccccc}
      \hline
      \noalign{\smallskip}
      Name & $\!$$N_{\rm Phot}$ & $N_{\rm H}^{\rm gal ~(2)}$ & 
         \multicolumn{4}{c}{free parameter fit} & 
         \multicolumn{3}{c}{fit with $N_{\rm H}$=$N_{\rm H}^{\rm gal}$} & 
         \multicolumn{3}{c}{fit with kT=500000 K} \\
          &                &                             &
           $N_{\rm H}$ & $kT$ & $\!\!$$Norm$$\!\!$ & $\chi^2_{\rm red}$ &
           $kT$ & $\!\!$$Norm$$\!\!$ & $\chi^2_{\rm red}$ &
           $N_{\rm H}$ & $\!\!$$Norm$$\!\!$ & $\chi^2_{\rm red}$ \\
     \noalign{\smallskip}
      \hline
     \noalign{\smallskip}
     PX And$\!$   &  137 & 0.387 & 2.50$\pm$1.10 & 180$\pm$180 & \,1.5 & 0.70 
                & {\bf 240$\pm$80} & {\bf \,0.51} &  0.71 
                & 32$\pm$8 & 2$\times$10$^9$ & 0.8 \\
     TT Ari   & 8955 & 0.619 & {\bf 0.12$\pm$0.02} & {\bf 325$\pm$15~} & {\bf 37.8} & 1.67 
                & 290$\pm$30 & 47.2 & 3.75 
                & $\!\!$2.2$\pm$0.1$\!\!$ & 5$\times$10$^4$ & 75 \\
     KR Aur   & 1105 & 4.042 & {\bf 0.19$^{\bf +0.40}_{\bf -0.09}$} & {\bf510$\pm$50~} & {\bf \,6.2} & 1.57
                & 285$\pm$20 & 16.1 & 2.8 
                & 32$\pm$2 & 2$\times$10$^{10}$ & 16 \\
     BZ Cam   &  737 & 0.788 & 1.2$^{+1.3}_{-1.0}$ & 270$\pm$60 &  10.5 & 1.15
                & {\bf 285$\pm$30} & {\bf 8.8} & 1.12
                & 35$\pm$2 & 8$\times$10$^{10}$ & 3.8 \\
     DW UMa$\!$   &   60 & 0.070 & 11.4$\pm$3.0 & 100$\pm$80 & 940 & 0.32 
                & 280$\pm$90 &  1.2 & 0.76 
                & 25$\pm$7 & 1$\times$10$^{9}$ & 1.6 \\
     MV Lyr   & 1397 & 0.575 & $\!${\bf 0.03$^{\bf +0.15}_{\bf -0.03}$}$\!\!$ & 
                  {\bf 490$\pm$50} & {\bf \,6.9} & 1.21 
                & 430$\pm$40 & 7.6 & 2.4 
                & 41$\pm$2 & 3$\times$10$^{11}$ & 24 \\
     V794 Aql$\!$ &   22 & 0.640 & 3.5$\pm$3.0 & 220$\pm$200 & 54.1 & 0.65
                & $\!\!${\bf 310$\pm$170}$\!\!$ & {\bf 16.8} & 0.58
                & 1.1$\pm$0.8 & 1610  & 2.3 \\
     VZ Scl   &   27 & 0.166 & 0.01$^{+1.2}_{-0.01}$ & 220$\pm$110 & \,0.6&1.75
                & $\!\!${\bf 210$\pm$110}$\!\!$ & {\bf 0.73} & 1.64
                & 1.2$\pm$0.9 & 308 & 2.0 \\
          \noalign{\smallskip}
      \hline
      \noalign{\smallskip}
  \end{tabular}

\noindent{\small $^{(1)}$ Note that photon statistics is very poor in several
                          cases, and fit results are an order of magnitude
                          estimate for $N_{\rm phot}<500$.  
                          $N_{\rm H}$ is in units of 10$^{21}$ cm$^{-2}$,
                          $kT$ in units of eV and 
                          $Norm$ in units of 10$^{-4}$ ph/cm$^2$/s. The fit
                          with the most reasonable values (weighting absorbing
                          column and luminosity versus reduced $\chi^2$) is 
                          printed 
                          bold (except for DW UMa -- see section 3.2.). \\
                 $^{(2)}$ Dickey \& Lockman (1990) }
   \label{fitres}
   \vspace*{-0.2cm}
   \end{table*}

\subsection{Optical data}

In order to evaluate the optical state of each VY Scl star during the time of 
the ROSAT X-ray observation all available brightness measurements have been 
collected. Most of the data, especially of the well sampled objects, were 
either supplied by the AAVSO (courtesy J. Mattei; PX And, BH Lyn, DW UMa, 
LX Ser, MVLyr) or taken from VSNET 
(http:/$\!$/www.kusastro.kyoto-u.ac.jp/vsnet/; PX And, DW UMa, V751 Cyg).
In some cases (PX And, BH Lyn and DW UMa) dense CCD observations during
the end of 1996 were included in the otherwise visual brightness estimates
obtained from AAVSO, and the faint phases of the eclipses of all these 
three objects have not been taken out, so that the apparent scatter is large.
Further data were supplied from Roboscope observations (courtesy K. Honeycutt),
the UK Schmidt plates (courtesy S. Tritton), and the Sonneberg Observatory
sky patrol (courtesy G. Richter). All these optical data are plotted in 
Fig. \ref{lc} together with the times of the
ROSAT observations. Since the main emphasis is on the correlation with 
ROSAT observations, no attempt has been made to maintain fine structures
in the individual light curves. Also, no color corrections have been
applied to account for measurements in different photometric bands
because these are usually rather small in VY Scl stars (typically 
B--V $\approx$0).

\section{Results}

\subsection{X-ray spectral analysis}

A comparison of the times of ROSAT observations with the optical state of the 
respective VY Scl star shows that in all cases of ROSAT PSPC observations 
(with the exception of 
DW UMa, see below) in which the optical state is known (80\%) it has been the 
high state during the ROSAT observation (Fig. \ref{lc}). 
Thus, in the following I will exclusively  talk about X-ray
emission of VY Scl stars during their optical high state.

All VY Scl stars detected with the ROSAT PSPC show a surprising
uniformity in their hardness ratios, i.e. their global spectral shape.
Spectral fits to the objects with 
more than 100 photons consistently result in the fact that power law,
thermal bremsstrahlung or Raymond-Smith models give worse results than 
blackbody fits (this has already been noted earlier by van Teeseling \etal\ 
1996 for two objects), and that the range of blackbody temperatures is rather 
narrow (between 0.25--0.5 keV). As an example, the best-fit of a thermal
brems\-strah\-lung spectrum to the high signal-to-noise ratio spectrum of 
TT Ari results in a reduced $\chi^2$=1.99, the best-fit Raymond-Smith thermal 
plasma model in a reduced $\chi^2$=1.87  and the best-fit power law model in 
a reduced $\chi^2$=2.23 as compared to the reduced $\chi^2$=1.67 of a 
blackbody model (see below  for an explanation why the reduced $\chi^2$ values 
are such high). Fig. \ref{cont} shows the significance contours in the 
$N_{\rm H}-kT$-plane for the 4 objects with $>$500 photons demonstrating
the rather tight constraints on both the temperature and luminosity.

\begin{figure*}
    \vspace*{-.8cm}
    \vbox{\psfig{figure=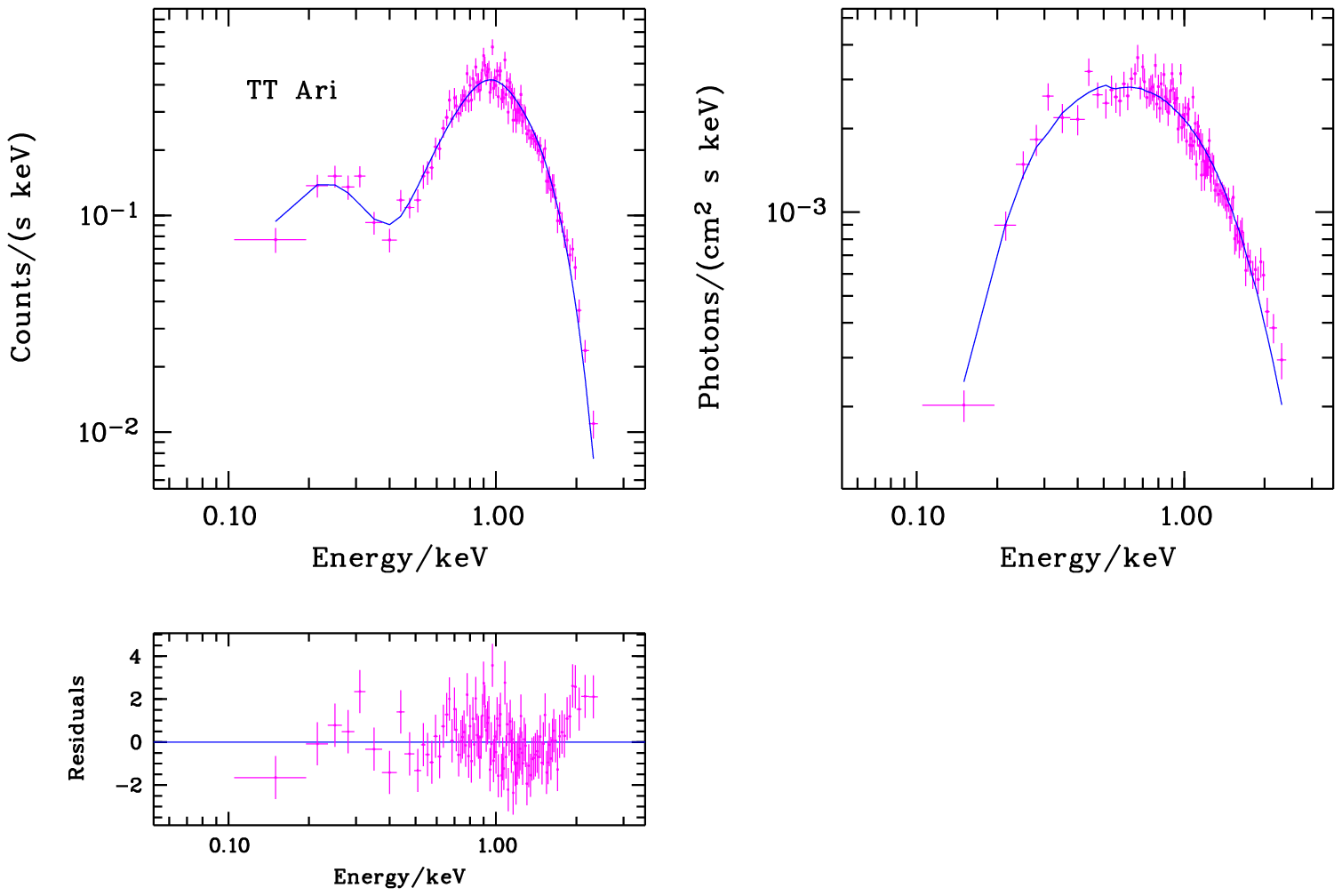,width=8.cm,%
           bbllx=1.5cm,bblly=1.3cm,bburx=9.2cm,bbury=11.6cm,clip=}}\par
    \vspace*{-10.6cm}\hspace*{8.3cm}
    \vbox{\psfig{figure=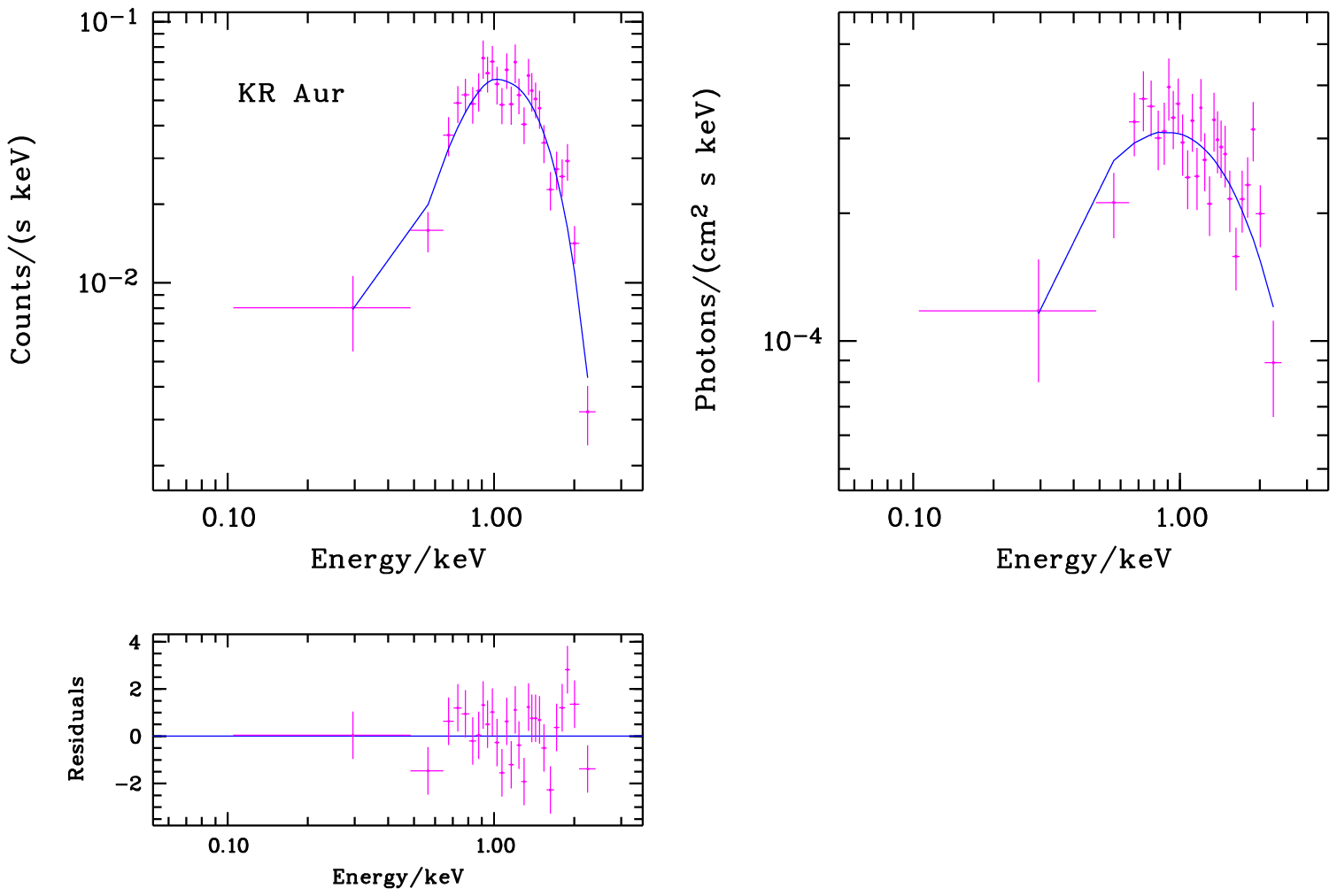,width=8.cm,%
           bbllx=1.5cm,bblly=1.3cm,bburx=9.2cm,bbury=11.6cm,clip=}}\par
    \vbox{\psfig{figure=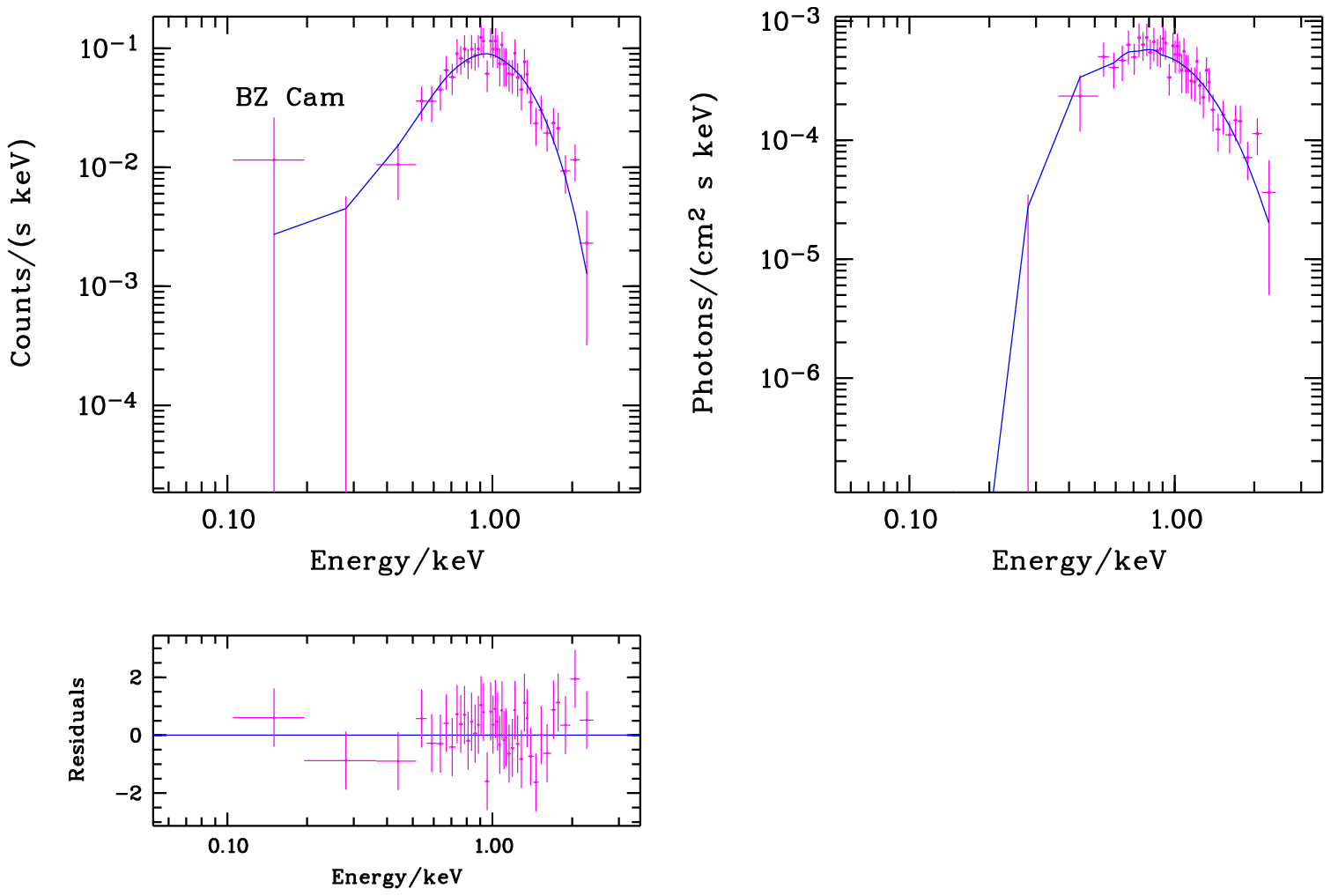,width=8.cm,%
           bbllx=1.5cm,bblly=1.3cm,bburx=9.2cm,bbury=11.6cm,clip=}}\par
    \vspace*{-10.6cm}\hspace*{8.3cm}
    \vbox{\psfig{figure=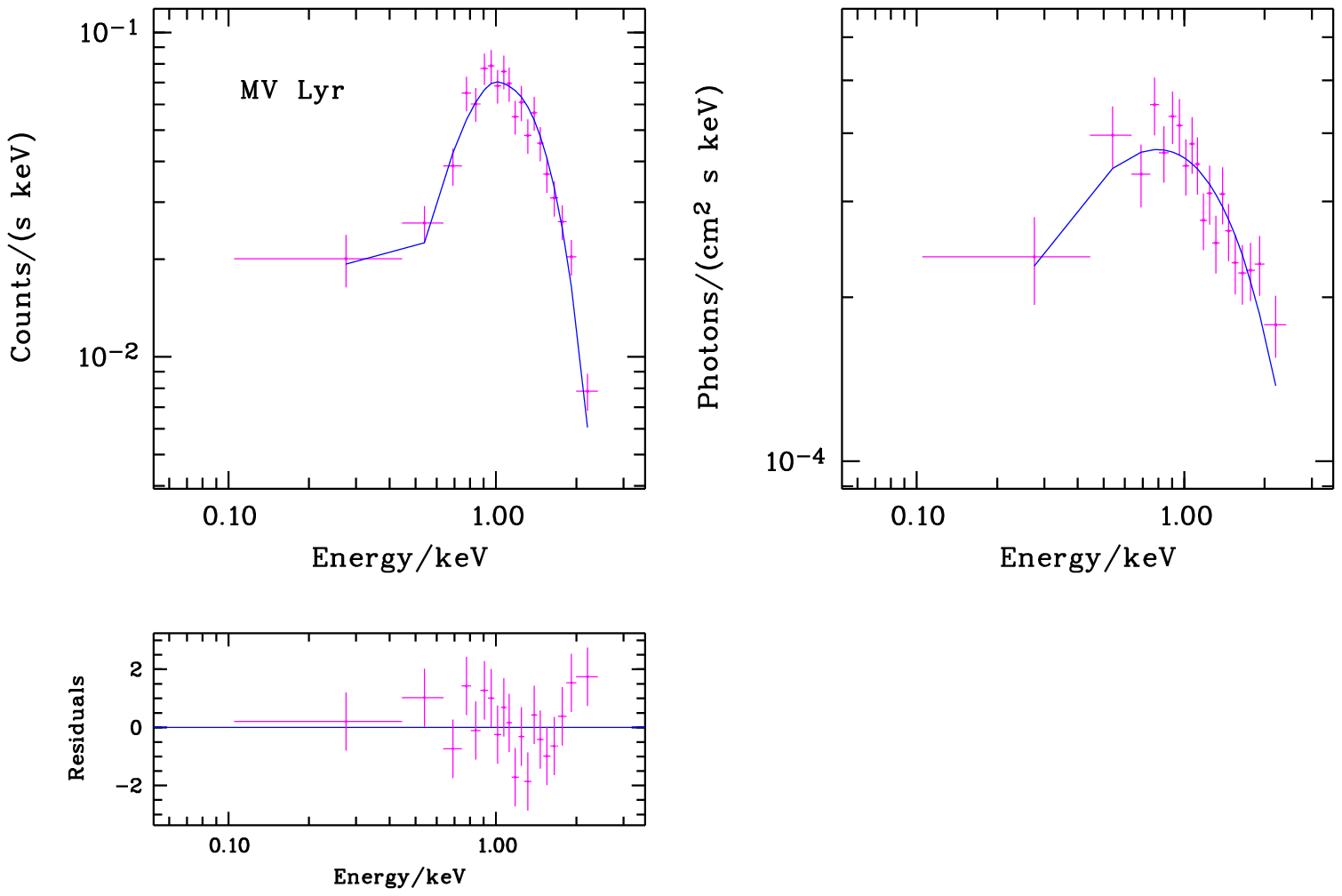,width=8.cm,%
           bbllx=1.5cm,bblly=1.3cm,bburx=9.2cm,bbury=11.6cm,clip=}}\par
    \caption[pha]{ROSAT PSPC X-ray spectra of the VY Scl stars
           TT Ari (Aug. 1991), KR Aur (Sep./Oct. 1992), BZ Cam (Sep. 1992)
           and MV Lyr (Nov. 1992). The top panels for each object show
           the count rate spectrum together with the model folded with the 
           PSPC detector response, while the lower panels show the deviation 
           between model and data (residuals) in units of $\sigma$.
           All spectra have been fitted with a 
           blackbody model and give temperatures between 0.25--0.5 keV.
           In some spectra the residuals are large indicating more complex
           spectra, possibly due to Fe L-shell or O K-shell emission lines.
          }
      \label{spec}
   \end{figure*}

\begin{figure*}
%    \vspace*{-.8cm}
    \vbox{\psfig{figure=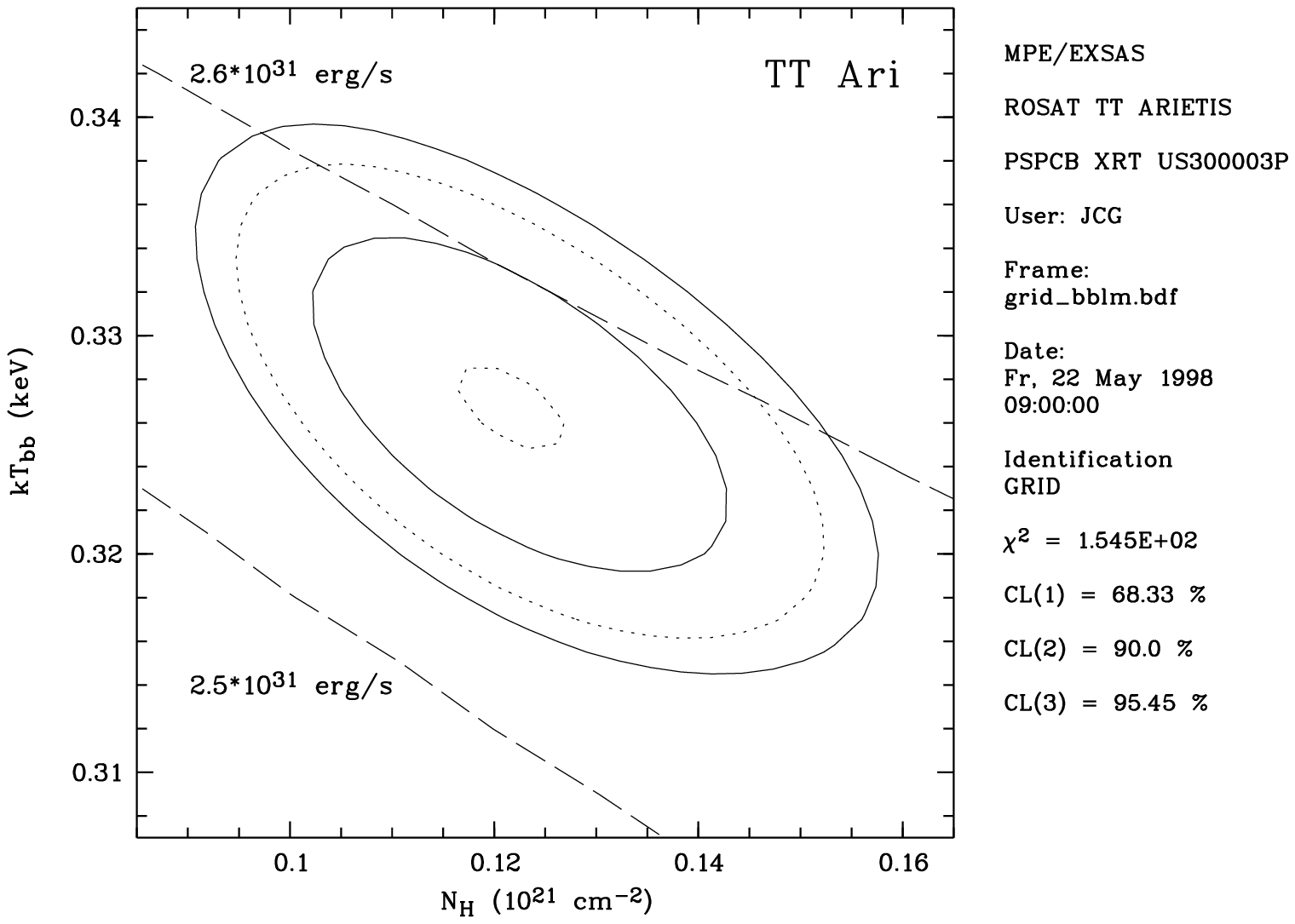,width=8.5cm,%
           bbllx=1.9cm,bblly=1.1cm,bburx=13.5cm,bbury=12.2cm,clip=}}\par
    \vspace*{-8.1cm}\hspace*{8.8cm}
    \vbox{\psfig{figure=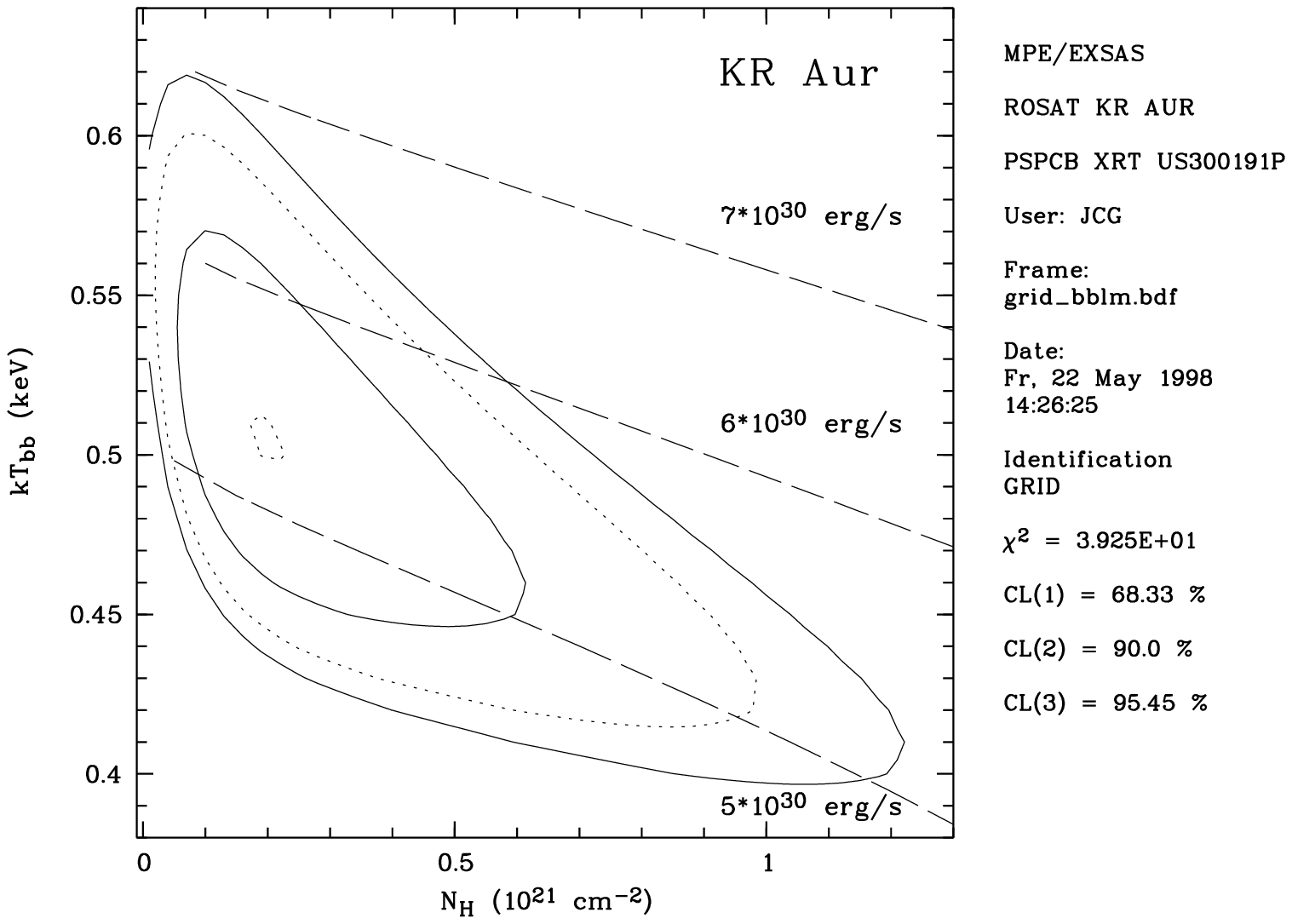,width=8.5cm,%
           bbllx=1.9cm,bblly=1.1cm,bburx=13.5cm,bbury=12.2cm,clip=}}\par
    \vbox{\psfig{figure=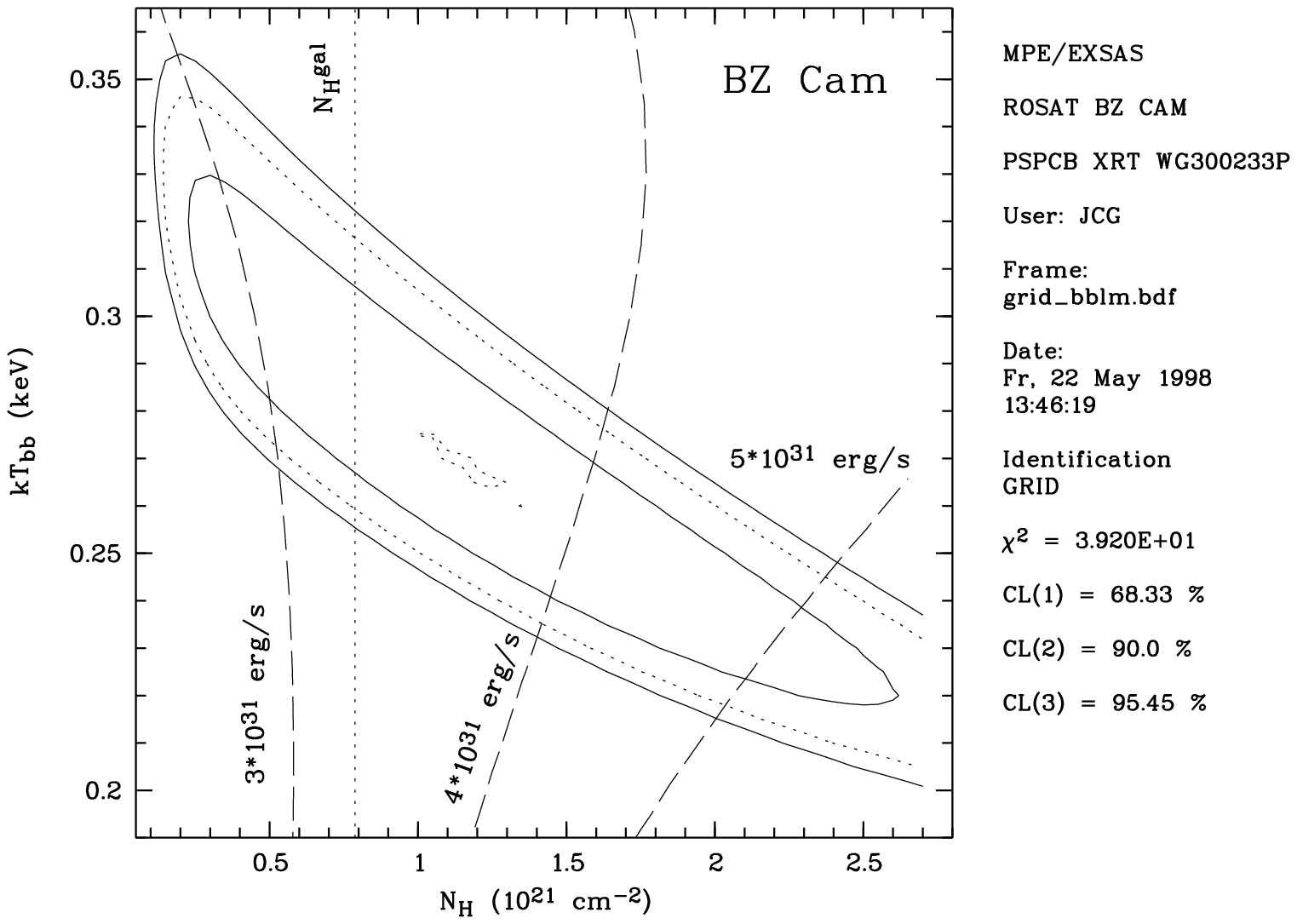,width=8.5cm,%
           bbllx=1.9cm,bblly=1.1cm,bburx=13.5cm,bbury=12.2cm,clip=}}\par
    \vspace*{-8.1cm}\hspace*{8.8cm}   
    \vbox{\psfig{figure=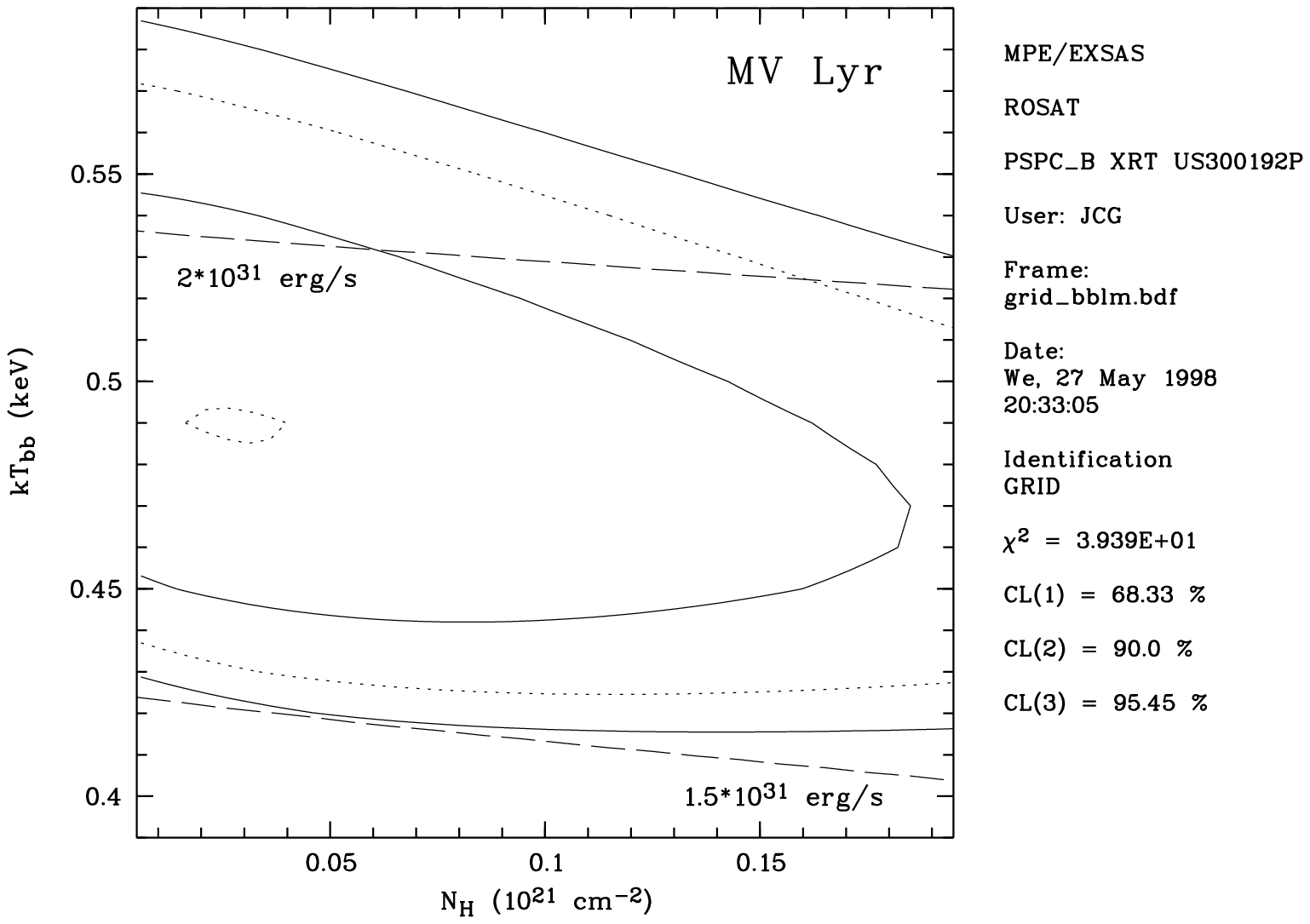,width=8.5cm,%
           bbllx=1.9cm,bblly=1.1cm,bburx=13.5cm,bbury=12.2cm,clip=}}\par
    \caption[cont]{The 50\%, 68\%, 90\% and 95\% significance contours
          of the blackbody model fits of the four sources with more than
         500 photons (see Fig. \ref{spec}) in the $N_{\rm H}-kT$ plane.
         Overplotted are the lines of constant luminosity (dashed), 
         demonstrating the rather tight constraints imposed by even 
         low-statistics spectra.
          }
      \label{cont}
   \end{figure*}

 As the high signal-to-noise spectra show (Fig. \ref{spec}), they deviate
systematically from a simple blackbody model (see the {\it bad} reduced 
$\chi^2$ values of the best-fits of TT Ari, KR Aur and MV Lyr in 
Tab. \ref{fitres} and the residuals in Fig. \ref{spec}). In the cases of MV Lyr
and KR Aur this has been noted already by Schlegel \& Singh (1995) who 
speculated about the presence of Fe L-shell emission. I will not elaborate on 
this any further, and only note that this could be a generic property of VY Scl
stars which has to be investigated with the next generation X-ray telescopes
with high sensitivity and spectral resolution.

%%-----------------------------------------------------------
%%______________________________________________________________

%%-----------------------------------------------------------

Given the poor knowledge of the distance and the effect of the absorbing 
column in a few systems it is worth noting that besides the spectral 
parameters also the X-ray luminosities are in a narrow range between
log L$_{\rm X}$ = 30.7--31.5 erg/s. (This is independent of the interpretation
as blackbody emission and has been noted also for other non-magnetic
cataclysmic variables by van Teeseling \etal\ 1996).
 A different notion of the same fact is
that the implied blackbody radii of the emitting region are within the
narrow range of 50--120 m.

\subsection{X-ray temporal analysis}

\begin{figure}
   \vbox{\psfig{figure=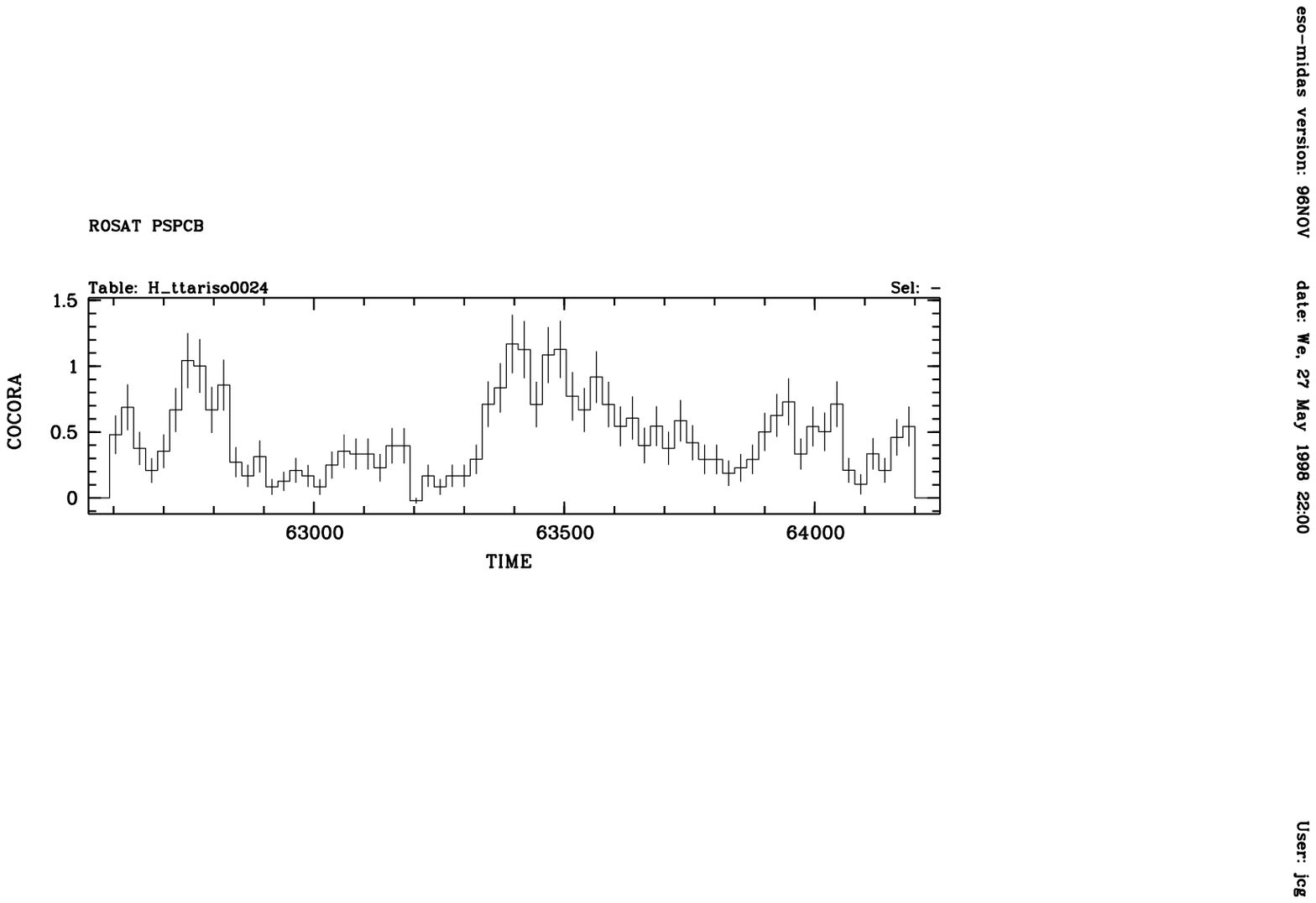,width=8.7cm,%
           bbllx=1.8cm,bblly=5.2cm,bburx=13.3cm,bbury=8.83cm,clip=}}\par
   \vbox{\psfig{figure=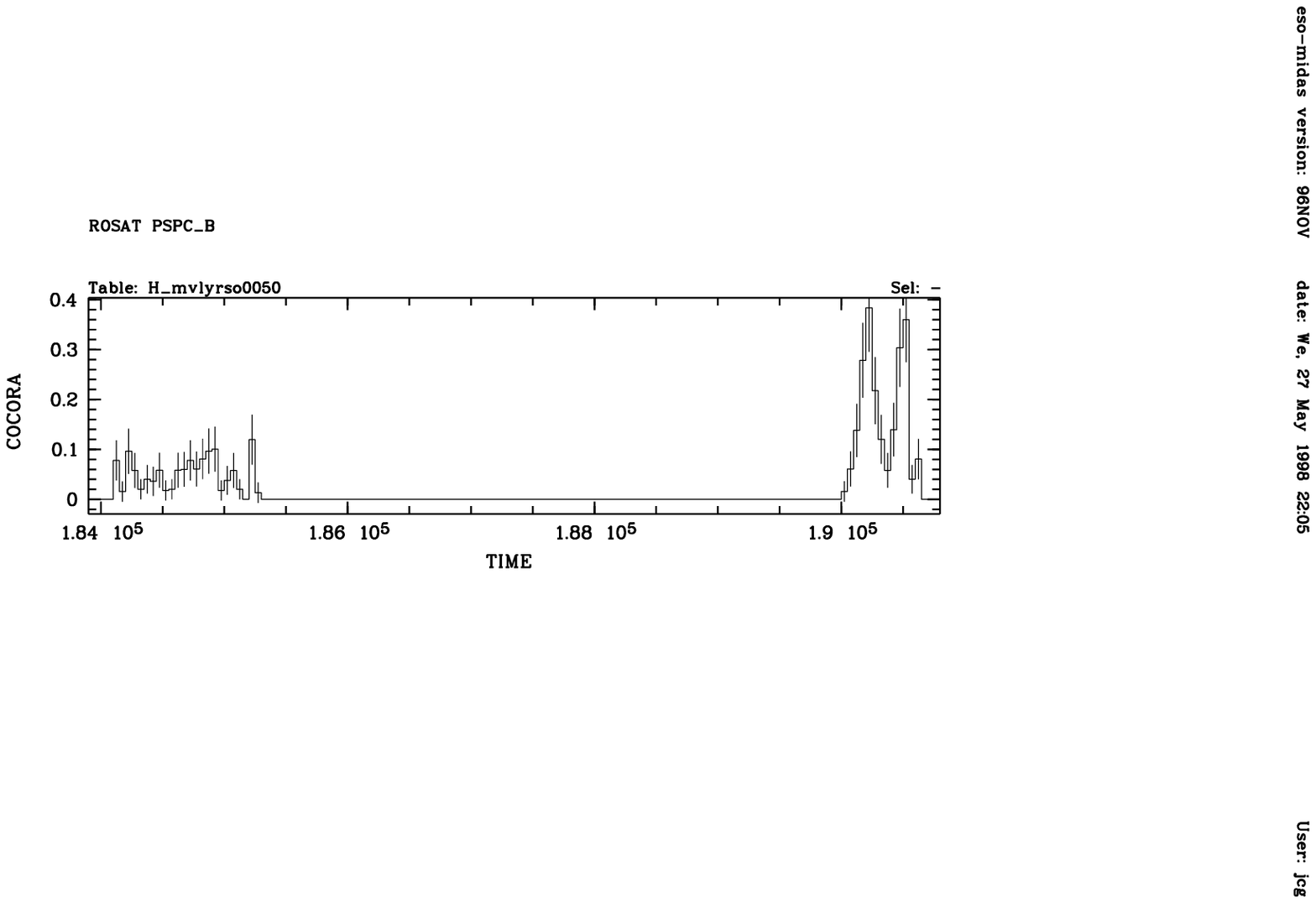,width=8.7cm,%
           bbllx=1.8cm,bblly=5.2cm,bburx=13.3cm,bbury=8.82cm,clip=}}\par
   \caption[lc]{Variation of the X-ray count rate of TT Ari (top; bin size of
    25 sec) and MV Lyr (bottom; bin size of 50 sec) with time as measured 
       with the ROSAT PSPC around their mean values of 0.36 cts/s (TT Ari) and
       0.07 cts/s (MV Lyr).
          }
      \label{indlc}
   \end{figure}

A detailed X-ray temporal analysis is only possible for the four brightest
objects, and is beyond the subject of this paper. For the purpose of
showing the degree of short-term variability two examples are given
in Fig. \ref{indlc} (TT Ari at the top and MV Lyr). It consists of two time
short stretches out of the total exposure and demonstrates that the
X-ray emission of TT Ari and MV Lyr are strongly variable by factors of
about 3--5 on time scales of 20--100 sec.

%%-----------------------------------------------------------

\subsection{Notes on individual objects}

\noindent{\it TT Ari} pointed observations with ROSAT have been investigated 
in detail by Baykal \etal\ (1995). My spectral fit result is in 
agreement with their findings (except for the normalization which seems to
be a misprint of their exponent given the nearly exact coincidence in figure).
In particular, as can be seen from their Tab. 1, they also find that the 
blackbody model gives the lowest reduced $\chi^2$ though they do not comment 
on this. Baykal \etal\ (1995) find in addition that the X-ray emission is 
strongly modulated with orbital period and, based on the phasing, relate 
the X-ray emission to an extended region around the impact site of the 
accretion stream onto the disk.

\noindent{\it BZ Cam} has been investigated already by van Teeseling \&
Verbunt (1994) and nothing new is reported here. Our results are in perfect
agreement in all respects.

\noindent{\it DW UMa} is the only source in the present sample which deviates 
substantially from the rather tight range of hardness ratios, i.e. being much
softer. Looking at the optical light curve during the PSPC observation 
shows that DW UMa was recovering from a minimum at the time of the 
X-ray observation. Unfortunately, the low number of counts does not allow 
to quantify this effect (Tab. \ref{fitres}).

\noindent{\it MV Lyr}:
The X-ray emission of MV Lyr during the optical high state is similar to
that of all other VY Scl stars described here. However, when I observed
MV Lyr on 28 May 1996 at the end of the 9 week
optical low state, no  X-rays were detected at all. The upper limit imposed
by this observation (based on the assumption of a hard X-ray source and 
thus a PSPC/HRI conversion factor of about 3) is a factor 30 lower than
the flux measured during the optical high state in 1990 and 1992.

\noindent{\it V751 Cyg} has by purpose been observed twice during its
recent optical low state. The detection of strong X-ray emission during these
observations and its interpretation using also additional archival IUE and
quasi-simultaneous optical observations will be given in a separate paper
(Grei\-ner \etal\ 1998).

\noindent{\it VY Scl} has been
 marginally detected (7 photons) at a position 50\asec\ off the nominal
optical position during the ROSAT all-sky survey. Given the few photons 
and the ROSAT all-sky scanning mode,
these photons could be related to VY Scl though the distance between X-ray 
and optical position for all other detections is less than 17\asec\ 
(see Tab. \ref{xsurvlog}). If the detection is true, VY Scl would be the second
brightest X-ray emitter among its class (based on its X-ray count rate), 
suggesting either a rather small
distance or a high intrinsic luminosity. While this can be verified in future
X-ray observations, I conservatively take the measured flux at 50\asec\ off 
the nominal position as an upper limit for VY Scl (see Tab. \ref{xsurvlog}).

\section{Discussion}

\subsection{Comparison between VY Scl stars and dwarf novae}

Not much is known on the systematics of X-ray behaviour of VY Scl stars.
It may therefore be helpful to compare the properties of VY Scl stars to other
non-magnetic cataclysmic variables, in particular dwarf novae.
The general picture
(of dwarf novae) is that the X-ray intensity follows the optical brightness 
though sometimes with considerable delays. At a closer look is appears that 
during quiescence dwarf novae are only detectable at hard (few keV) X-rays 
while at outburst an additional strong soft component emerges. The standard
interpretation is that during quiescence a hot and optically thin boundary 
layer produces hard X-rays while during outburst the boundary layer becomes
optically thick thus producing the soft component.

The nature of the X-ray emission in dwarf novae during quiescence has been
demonstrated by using X-ray eclipse measurements to arise very close to 
the white dwarf (Mukai \etal\ 1997, van Teeseling 1997).
The lack of correlation between X-ray temperature or luminosity
with accretion rate (van Teeseling \etal\ 1996) and the evidence of orbital
modulation of the X-ray intensity from some non-eclipsing systems
(van Teeseling \& Verbunt 1994, Baykal \etal\ 1996) argue against the simple
boundary layer picture. On the other hand, the evidence for an 
anti-correlation between observable emission measure and orbital inclination
suggests that the X-rays are emitted close to the white dwarf and not by an 
optically thin cloud much larger than the white dwarf radius (van Teeseling 
\etal\ 1996).

Given the high accretion rate in VY Scl stars during optical high state one 
would naively expect that their X-ray spectra should be softer than those of
dwarf novae during quiescence and similar to dwarf novae in outburst.
This is not born out by observations. Dwarf novae during quiescence have
typically softer hardness ratios than high accretion rate systems including
VY Scl stars (Patterson \& Raymond 1985, van Teeseling \etal\ 1996). Moreover,
the finding that VY Scl stars during their optical high state are 
consistently described by blackbody models with temperatures in the narrow 
range of 250--500 eV extends an earlier finding of van Teeseling \etal\ (1996) 
that VY Scl stars and non-SU UMa systems either are intrinsically absorbed or 
do not have simple bremsstrahlung spectra.

These differences in X-ray behaviour between dwarf novae during outburst and 
VY Scl stars are not surprising because there are several other distinct 
differences other than at X-rays. These can be 
summarized as follows (Hack \& la Dous 1993, Warner 1995): 
\begin{itemize}
\item The rise and decline times are not similar. In dwarf novae a surprisingly
tight correlation was found between the decline time from outburst and the 
orbital period (Bailey 1975), suggesting that the size of the disk is 
somehow related to its cooling time after outburst. In VY Scl stars the rise
and decay times are much longer than this relation would imply.
\item In contrast to dwarf nova there are no traces of outburst behaviour
during the long low-state behaviour of VY Scl stars.
\item In many VY Scl stars the hump in the orbital light curve (usually
interpreted as being due to the hot spot where the accretion stream hits the 
disk) has a significantly shorter length ($\approx$0.3 $P_{\rm orb}$) than in 
dwarf novae ($\approx$0.5 $P_{\rm orb}$). It has been argued that this is no
physical difference but just due to the smaller intensity contrast between 
hot spot and disk continuum emission in VY Scl stars. However, in some
VY Scl stars the hump occurs after eclipse which has never been observed in
dwarf novae.
\item Flickering may cease during the optical low states in VY Scl stars 
whereas this has never been found in dwarf novae.
\item The optical/UV spectra of VY Scl stars show signs of higher excitation 
than in dwarf novae, most notably He\II\ 4686 and C\III/N\III\ 4650.
\item The UV spectra of VY Scl stars are quite different from those of dwarf
novae, the majority of them showing even cooler continua than in dwarf novae.
It has been suggested that this may be caused by a systematic difference in 
the masses of the white dwarfs.
\item The transitions between high and low states manifest themselves in
completely opposite color changes: in dwarf novae the flux changes are 
largest in the UV and only marginal at IR wavelengths whereas in VY Scl stars
mostly the optical and IR emission changes.
\end{itemize}

\noindent Given these differences between VY Scl stars and dwarf novae and the 
systematic difference of the X-ray spectra of VY Scl stars from those of dwarf 
novae in outburst, one may ask how the generally discussed X-ray emission 
regions in non-magnetic cataclysmic variables compare with the X-ray spectral 
characteristics of VY Scl stars. This will be discussed in turn in the 
following sections.

\subsection{Is a blackbody model reasonable?}

As noted in section 3.1. a blackbody model gives the lowest reduced $\chi^2$
values among the single component models tested. In the blackbody model
interpretation the combination of best-fit parameters and distances results
in an estimate of the size of the emitting area of only 50--120 m. This 
inferred size is much smaller than anything we know of in a cataclysmic
binary system. One therefore may be inclined to doubt the fit results.

However, as Fig. \ref{cont} shows the cross-relation between temperature and 
absorbing  column is such that the variations compensate in the sense that the
resulting luminosity is in a very narrow range (see the overplotted lines
of constant luminosity in Fig. \ref{cont}). This is due to the fact that
the maximum of the blackbody emission is well within the energy range of the
ROSAT PSPC. Thus, even with the larger uncertainties in the fit parameters
of the sources with $<$500 photons the luminosity is still rather well
constrained. Therefore, the uncertainties due to low photon numbers imply 
only a small uncertainty in the size of the emitting area.

One is therefore left to either adopt a blackbody model (with the parameters
as given in section 3.1.) or to reject a blackbody model interpretation at all.
The latter option finds additional support in the strength and location of the
fit residuals around $\approx$1 keV mentioned earlier. Schlegel and Singh
(1995) have suggested Fe L-shell emission as a possible explanation of these
residuals. Oxygen K-shell emission would be an at least similarly appropriate
alternative. If line emission is indeed present then the best continuum
model to compare with would actually be a hot plasma Raymond-Smith model.
As noted earlier, a Raymond-Smith model with solar abundances gives a poorer
fit to the ROSAT data than a blackbody model. While it maybe possible
to achieve reasonably low reduced $\chi^2$ values by tuning the relative
metal abundances or by introducing multi-temperature components, the 
anticipated results will be quite uncertain due to the given energy resolution
of the PSPC. Thus, I leave such an exercise for a future investigation
and continue with the discussion of possible emission sites.

\subsection{The classical boundary layer picture}

The high \mdot\ and the optical spectra of VY Scl stars during the optical
high state imply an optically thick accretion disk. According to standard
theory the accretion disk dominates the optical-UV-X-ray luminosity in most 
non-magnetic cataclysmic variables. That is, the observed features mostly
relate to the disk properties which in turn are governed by the mass transfer
rate \mdot.

A very early prediction of non-magnetic cataclysmic variable theory was for 
the existence of a hot boundary layer (Pringle 1977, Pringle \& Savonije 1979, 
Tylenda 1981). The matter at the inner edge of the 
accretion disk rotates with $\gax$1000 km/s (Keplerian) and will shear on to
the surface of the slowly rotating white dwarf (assumed to rotate below its 
break-up rate) thus dissipating its kinetic energy in the so-called
boundary layer (Lynden-Bell \& Pringle 1974). 
Thus, the boundary layer luminosity should equal that of the disk,
i.e. GM$_{\rm WD}$\mdot$_{\rm acc}$/2R$_{\rm WD}$ $\approx$ 
10$^{34}$--10$^{35}$ erg/s with M$_{\rm WD}$ and
R$_{\rm WD}$ being the mass and radius of the white dwarf, respectively and 
\mdot$_{\rm acc}$ being the mass accretion rate.
Depending on the accretion rate the boundary layer is expected to be either
optically thin (at low \mdot) with temperatures of $\approx$10$^8$ K 
(Pringle \& Savonije 1979, Tylenda 1981) or to form an optically thick ring
around the white dwarf equator (at high \mdot) with temperatures of 
200000--500000 K (Pringle 1977, Kley 1989).

This picture of an optically thick boundary layer has been elaborated 
in more detail by Popham \& Narayan (1995) by exploring the parameter
space of temperatures and luminosities depending on the mass and rotation of
the white dwarf, the accretion rate and the viscosity parameter.
Their predictions are in line with the earlier estimates and yield
maximum temperatures of about 800000 K and X-ray luminosities always larger
than 10$^{34}$ erg/s.

Thus, the observed X-ray characteristics of VY Scl stars (Tab. \ref{fitres})
during the optical high state with temperatures in the 3--5$\times$10$^6$ K
range and luminosities of 10$^{31}-10^{32}$ erg/s are distinctly different from
those expected in the classical boundary layer picture in that the
temperature is intermediate between the two types of boundary layer models
and the luminosity is lower than either type of model.

\subsection{Heavily absorbed classical boundary layer emission}

The UV observations with IUE over the last 2 decades have accumulated
much evidence that high accretion rate cataclysmic variables possess
strong stellar winds as revealed by the P Cygni profiles of UV resonance 
lines, predominantly C\IV 1549, N\V 1240 and Si\IV 1397.
It has consequently been argued that the lack of detecting the expected
luminous and soft boundary layer emission is due to the fact that
the wind is strong enough to either absorb the boundary layer emission 
considerably or even be optically thick for soft X-rays.

In the first case, one therefore would expect a heavily absorbed 200000--500000
K emission. I have tested this possibility by fitting the X-ray spectra
with a fixed blackbody temperature of $kT=43$ eV (500000 K).
Though it is difficult to distinguish for some low signal-to-noise spectra an 
unabsorbed high-temperature model from an absorbed low-temperature model, 
it can be seen from Tab. \ref{fitres} (last column) that applying this model 
generally results in much worse reduced $\chi^2$. The fits are generally so 
bad that it seems unlikely that this could be affected
by the systematic deviations from a smooth spectrum around 1 keV (as discussed 
above). For lower temperatures the fits get even worse. I tentatively 
conclude that a heavily absorbed classical boundary layer emission is not a 
promising explanation of the observed X-ray spectrum of VY Scl stars.

\subsection{X-ray emission from an optically thick wind}

If, as explained in the previous section, the wind is completely optically 
thick for the boundary layer emission, one would only see the thermal
emission from the wind itself (Mauche \& Raymond 1987, Hoare \& Drew 1993).
However, its size is expected to be much larger than the derived value of
50--120 m, and secondly 
its temperature is not expected to be considerably higher than
that of the boundary layer, and therefore thermal emission from the wind is 
not a likely candidate for the observed 0.2--0.5 keV emission in VY Scl stars.
Independent of the blackbody model interpretation the seemingly periodic
variation of the X-ray intensity in TT Ari (Baykal \etal\ 1995) as well as
the short-term variability on 20--100 sec scales argues 
against a wind interpretation.

Alternatively, if the winds are radiation driven then one may expect that 
shocks will be present in the wind. This could give rise to hard X-ray emission
similar to the one observed in the winds of O stars 
(Bergh\"ofer \& Schmitt 1994).

\subsection{Coronal region above the accretion disk}

There is extensive observational evidence for the existence of hot, 
low-density coronae above accretion disks. Various processes may lead 
to the formation and heating of such a corona, e.g. X-ray illumination of the
disk (Begelman \etal\ 1983), thermal instabilities in the disk atmosphere
due to the density dependence of the opacity (Shaviv \& Wehrse 1986) or 
the vertical propagation of sound waves in the disk (Murray \& Liu 1992).
The common feature of all these models is that the emission is optically
thin and extended to approximately the size of the binary system.
The evidence of superior spectral fits of blackbody models as opposed
to optically thin emission argues against a hot corona as the origin
of the 0.25-0.5 keV emission in VY Scl stars during their optical
high state. As noted in the previous subsection, independent of the blackbody 
model interpretation the seemingly periodic
variation of the X-ray intensity in TT Ari (Baykal \etal\ 1995) as well as
the short-term variability on 20--100 sec scales argues
against a large emission area interpretation.

\subsection{Coronal emission of the late-type companion star}

For a few of the VY Scl stars the spectral type of the companion has been 
determined. Given the small range in orbital period of the VY Scl stars
the small range in spectral type of M3--M5 is not unexpected, and also
coincides with the relation between period and secondary spectral type
found for other cataclysmic variables (Echevarria 1983).
Typical X-ray luminosities of the coronal emission of M dwarfs 
(Fleming \etal\ 1993, Pye \etal\ 1994) range between
10$^{27}$--10$^{28}$ erg/s. Though the secondaries in cataclysmic variables
are expected to be rapidly rotating and magnetically active, and thus their 
X-ray emission could be somewhat larger than that of field stars, it seems
unlikely that it would be larger by an order of magnitude or more.
Also, the temperature of the observed X-ray emission of late-type field stars
are of order a factor 10 lower than those presented here for VY Scl stars
during optical high state. Finally, the drop of the X-ray emission from MV Lyr
during the optical low state by a factor of 5--7 as compared to the optical 
high state emission would be difficult to reconcile with coronal emission from 
the secondary. Thus, the secondaries in VY Scl stars are not expected to be 
responsible for the observed X-ray emission during optical high state.

\subsection{Evaporation of the accretion disk}

The disk instability model successfully explains dwarf novae outbursts except
of two observational features, namely the UV lag behind the optical rise and
the appearance and gradual decline of hard (few keV) X-ray emission during 
the optical decline after an outburst. Meyer \& Meyer-Hofmeister (1994)
have shown that above a cool disk a coronal siphon flow exists which evaporates
mass from the disk into the corona and accretes it onto the white dwarf.
While during outburst the X-ray radiation of dwarf novae is thought to
originate from shocks connected with radiation driven winds, X-rays during
quiescence are caused by either the white dwarf boundary layer (higher 
temperature and thus harder X-rays) or the corona in connection with the 
evaporation process (lower temperature, i.e. softer X-ray emission).
The temperature in the corona scales as 
$T \approx 10^{7.6} {M/{\rm M_\odot} \over r_{9.5}}$ K 
(with r$_{9.5} = r / 10^{9.5}$ cm) and thus typically is in the few keV range
(Meyer \& Meyer-Hofmeister 1994). While this scenario produces a long delay
between X-ray and optical outburst emission (as observed) and also explains 
the change
in the X-ray spectrum along the outburst of dwarf novae, it is not clear 
whether or not the whole picture is applicable also to VY Scl stars which spend
most of their time in the high state and thus during this state should have
appreciable accretion via the corona. Estimates of the spherical accretion 
onto the white dwarf via the corona range around 10$^{-11}$ \msun/yr
(Meyer \& Meyer-Hofmeister 1994) as opposed to the 10$^{-8}$ \msun/yr
deduced for VY Scl stars during their optical high states.
Also, the size of the corona is much larger than the deduced 50--120 m 
emission size from the blackbody spectra.

\subsection{Stream impact region}

The region where the stream impacts on the outer rim of the accretion
disk could have a size which is comparable to that derived from the
blackbody fits. Also, since the accretion rate in VY Scl stars is high,
the stream impact is expected to have substantial effects on the local
dynamics, e.g. heating up the impact point, overflowing the outer disk
or significant splashing leading to expansion of material in all directions
(Armitage \& Livio 1998, Hellier 1998, Spruit \& Rutten 1998).
However, both observations and modelling consistently result in 
temperatures of 100000 K at maximum, a factor of more then 10 below the
values derived from the ROSAT PSPC spectral fits.

\section{Conclusions}

The X-ray emission of VY Scl stars during optical high states is 
characterized by 0.25--0.5 keV blackbody emission from a 50--120 m
sized region. In the case of MV Lyr I found evidence for a drop in 
X-ray flux by a factor of $>$5--7 during an optical low state observation
suggesting that the emission is related to the change in accretion 
rate (which is thought to accompany the high/low state transitions) 
and not to the secondary coronal emission.
At this time, I cannot offer a simple explanation for this
emission. I cannot exclude the possibility that with the spectral
resolution of the ROSAT PSPC the X-ray continuum determination is
in error, and one may be fooled by emission which is not optically thick.
Future X-ray observations with higher spectral resolution are thus clearly
demanded.

It seems worth investigating the properties of other nova-like cataclysmic 
variables which do not belong to the VY Scl class to check whether the 
X-ray properties found for VY Scl stars are unique to this sub-class or are 
related to the high accretion rate in nova-like CVs in general.

\begin{acknowledgements}
I'm grateful to R. Di\thinspace Stefano for very fruitful discussions
and thank R. Popham, A. van Teeseling and S. Komossa for helpful comments
on an earlier version of this paper.
Much of the optical data presented in Fig. \ref{lc} were taken from the VSNET.
Additional data were provided by Janet Mattei (AAVSO), Kent Honeycutt
(Indiana Univ.), Sue Tritton (UK Schmidt plates), and Gerold Richter 
(Sonneberg Observatory) which I greatfully acknowledge. 
JG is supported by the German Bundesmi\-ni\-sterium f\"ur Bildung, 
Wissenschaft, Forschung und Technologie (BMBF/DLR) under contract No. 
FKZ 50 QQ 9602 3. The \ros\, project is supported by BMBF/DLR and the
Max-Planck-Society. This research has made use of the Simbad database, 
operated at CDS, Strasbourg, France. 
\end{acknowledgements}

\end{document}